\documentclass[useAMS,usenatbib]{mn2e}

\usepackage{graphicx}

\usepackage{amsmath,amssymb,amsfonts} 

\defcitealias{Dibietal2012}{D12}

\title[Spectra from radiative GRMHD simulations of Sgr A*]{Self-consistent spectra from radiative GRMHD simulations of accretion onto Sgr A*}
\author[S. Drappeau et al.]{S. Drappeau,$^{1}$\thanks{E-mail:
s.drappeau@uva.nl} S. Dibi,$^{1}$ J. Dexter,$^{2}$ S. Markoff,$^{1}$ and P.C. Fragile$^{3}$\thanks{KITP Visiting Scholar, Kavli Institute for Theoretical Physics, Santa Barbara, CA}\\
$^{1}$Astronomical Institute ``Anton Pannekoek", University of Amsterdam, Postbus 94249, 1090 GE Amsterdam, The Netherlands\\
$^{2}$Theoretical Astrophysics Center and Department of Astronomy, University of California, Berkeley, CA 94720-3411, United States\\
$^{3}$Department of Physics and Astronomy, College of Charleston, 66 George St., Charleston SC 29424, United States}
\begin{document}

\date{Accepted 2012 ?? ??. Received 2012 ?? ??; in original form 2012 ?? ??}

\pagerange{\pageref{firstpage}--\pageref{lastpage}} \pubyear{2012}

\maketitle

\label{firstpage}

\begin{abstract}
  We present the first spectral energy distributions produced self-consistently by 2.5D general relativistic magneto-hydrodynamical (GRMHD) numerical simulations, where radiative cooling is included in the dynamical calculation.  As a case study, we focus on the accretion flow around the supermassive black hole in the Galactic Centre, Sagittarius A* (Sgr A*), which has the best constrained physical parameters. We compare the simulated spectra to the observational data of Sgr A* and explore the parameter space of our model to determine the effect of changing the initial magnetic field configuration, ion to electron temperature ratio $T_i/T_e$ and the target accretion rate. We find the best description of the data for a mass accretion rate of $\sim 10^{-9} M_{\odot}$/yr, and  rapid spin ($0.7 < a_* < 0.9$). The submillimeter peak flux seems largely independent of initial conditions, while the higher energies can be very sensitive to the initial magnetic field configuration. Finally, we also discuss flaring features observed in some simulations, that may be due to artifacts of the 2D configuration.
\end{abstract}


\section{Introduction}
The best studied low-luminosity active galactic nucleus (LLAGN) is the supermassive black hole in the centre of the Milky Way, Sgr A*, discovered originally via its strong radio continuum emission by \citet{BalickBrown1974}. Sgr A* is a unique system because of its proximity compared to the centres of other galaxies, and a multitude of intensive single- and multi-wavelength campaigns having been conducted from the radio through high-energy $\gamma$-rays over the last decades (see references in \citealt{MeliaFalcke2001} and \citealt{Genzeletal2010}).  These studies provide remarkably stringent constraints on Sgr A*'s properties. The current best mass, distance and mass accretion rate values are $M = 4.3 \pm 0.5 \times 10^{6} M_{\odot}$, $D = 8.3 \pm 0.4$ kpc and $2 \times 10^{-9} M_{\odot} \, \mathrm{yr^{-1}} < \dot{M} < 2 \times 10^{-7} M_{\odot} \, \mathrm{yr^{-1}}$, respectively (\citealp{Reid1993, Ghezetal2008, Gillessenetal2009a, Gillessenetal2009b, Boweretal2005, Marroneetal2007}).

The above constraints make Sgr A* the perfect candidate to test theoretical models of the accretion processes at low accretion rates.  In particular, high precision data from close to the event horizon allow us to study the detailed physics of the accretion flow and potential jet launching, in the extreme regime where gravity is important. Furthermore, because Sgr A* is representative of the majority of SMBHs today, often lurking below the detection threshold of even our most sensitive telescopes, we can use it to get a handle on the contribution from these very weak AGN to their host galaxies.

In the attempt to understand the nature of accretion flows, many semi-analytical models have been developed. Accretion disc models like the thin disc developed by \citet{ShakuraSunyaev1973}, advection-dominated accretion flow \citep[ADAF;][]{NarayanYi1994}, advection-dominated inflow-outflow solutions \citep[ADIOS;][]{BlandfordBegelman1999}, convection-dominated accretion flow \citep[CDAF;][]{QuataertGruzinov2000a} or Bondi accretion \citep{Melia1992, Melia1994} have been successful at fitting data from many sources. However, by nature of the semi-analytic approach which cannot model turbulence, none of these models can accurately address the role played by the magnetic field in the dynamics of accretion disc via the magnetorotational instability \citep[MRI;][]{BalbusHawley1991}.

Recent breakthroughs in parallel numerical simulations now allow extensive fluid simulations with full general relativity (GR) and magneto-hydrodynamic (MHD) treatment in reasonable timescales.  Such simulations have finally provided the opportunity to study the dynamical properties of an entire system in a complementary manner to the semi-analytical models. In particular, various GRMHD codes have been employed by several groups over the past few years, in order to perform detailed theoretical studies of the accretion flow around Sgr A* \citep[e.g.][]{Dexteretal2009, Moscibrodzkaetal2009, Dexteretal2010, Hilburnetal2010, Shcherbakovetal2010, Moscibrodzkaetal2011, DexterFragile2012, Dolenceetal2012}, with the goal of reproducing Sgr A*'s spectra. These studies all share a common approach in which the radiative losses are not included in the simulations themselves, but rather first a dynamical model is calculated in GRMHD, and then the final outputs are fed into a separate post-processing routine to calculate the resultant spectrum. These studies justified ignoring the inclusion of cooling because Sgr A* is so underluminous that radiative losses are likely not strong enough to affect the dynamics of the system.

In a companion paper (\citealp{Dibietal2012}, hereafter \citetalias{Dibietal2012}), we assess, for the first time, the importance of the radiative cooling in numerical simulations of Sgr A* by using \texttt{Cosmos++}, an astronomical fluid dynamics code that takes into account radiative losses self-consistently in the dynamics \citep{Anninosetal2005,Fragileetal2012}. We show that, for Sgr A*, cooling effects on dynamics can indeed be neglected. However, the effects of cooling at higher accretion rates (relevant for most nearby LLAGN) are not negligible.

In this paper, we describe the implementation and results from the cooling routines used in the simulations of Sgr A* presented in \citetalias{Dibietal2012}, and present the first self-consistently calculated spectra in order to explore the new parameter space.  We examine the influence the spin and the initial magnetic field configuration have on the simulated spectra, and compare to the previous non-cooled calculations. Although we find that self-consistent treatment of radiative losses is not important for the case of Sgr A*, we demonstrate that it will be for most nearby LLAGN.

Section~\ref{sec:obs} describes the observational constraints on Sgr A*. In Section~\ref{sec:methods}, we present the derivation of emissivity expressions that generate the cooling rates employed in \texttt{Cosmos++}. We also discuss numerical limitations of our simulations, as well as the assumptions made when generating the spectra. In Section~\ref{sec:results}, we present the results obtained by comparing our spectra to observational data, and in Section~\ref{sec:discussion} we discuss them. Finally, in Section~\ref{sec:summary}, we summarise our conclusions and suggest future improvements.

\section{Observational constraints}
\label{sec:obs}
The radio spectrum of Sgr A* \citep{Serabynetal1997,Falckeetal1998,Anetal2005} shows a slight change in the spectral index above 10 GHz \citep{Falckeetal1998}, peaking in a so-called \textit{submillimeter bump}. \citet{Aitkenetal2000} reported the detection of linear polarisation from this submm bump, that is not observed at longer wavelengths. This change implies that the radio emission and the submm bump originate from distinct but contiguous regions in the system. Very long baseline interferometry measurements have limited the size of the submm-emitting region to be $\simeq 4$ Schwarzschild radii \citep{Doelemanetal2008}.

Limited angular resolution and sensitivity in the far- and mid-infrared bands makes it impossible to distinguish between the emission from Sgr A* and the surrounding sources, resulting in no observations in the far-infrared band and only upper-limits on the flux in the mid-infrared \citep{Schodeletal2011}. Detections in the near-infrared \citep[e.g.,][]{Davidsonetal1992,Herbstetal1993,Stolovyetal1996,Telescoetal1996,Mentenetal1997,Schodeletal2007,Schodeletal2011} and in the X-ray \citep{Baganoffetal2003,Belangeretal2006}, show a quiescent state lower in flux than the submm bump, where the radiation power peaks. However, a few times a day, Sgr A* experiences rapid increases in the near-infrared flux \citep{Hornsteinetal2002,Genzeletal2003,Ghezetal2004,Eckartetal2006,Eckartetal2008,YusefZadehetal2008,DoddsEdenetal2011,Hauboisetal2012}, where brighter flares \citep[$>10$ mJy;][]{DoddsEdenetal2011} are often associated with simultaneous X-ray flares \citep{Baganoffetal2001,Goldwurmetal2003,Porquetetal2003,Belangeretal2005,Porquetetal2008}.

These multi-wavelength observations provide tight constraints on the physics of Sgr A*. \citet{ShakuraSunyaev1973} type radiatively efficient, thin disc models are excluded as they predict an observed infrared flux several orders of magnitude higher than the upper-limits obtained \citep{FalckeMelia1997}. The presence of linear polarization and the constraints on the Faraday rotation currently limit the mass accretion rate of Sgr A* to be much smaller than the Bondi accretion rate, of the order of $\sim 10^{-8} M_{\odot}$ \citep{Aitkenetal2000,Boweretal2003,Marroneetal2007}. Such a low accretion rate in fact excludes the ``classical'' ADAF \citep{Narayanetal1998} and Bondi \citep{Melia1992} accretion models for Sgr A*, as these models invoke higher accretion rates \citep[see, e.g.,][]{Agol2000,QuataertGruzinov2000b}.  In the meantime, many other models have been developed that are still consistent with the current limits. Radiatively inefficient accretion flow models \citep[RIAF;][]{BlandfordBegelman1999,QuataertGruzinov2000a,Yuanetal2003} argue that the submm emission is produced via synchrotron radiation from a thermal distribution of electrons, in the innermost region of the accretion flow, which could also be synonymous with the base of the jets \citep{FalckeMarkoff2000,Yuanetal2002}. This synchrotron emission is then inverse Compton upscattered by these same electrons, resulting in a second peak that contributes to the X-ray emission during flares. The radio emission can originate from either a non-thermal tail of electrons produced in a RIAF \citep{Yuanetal2003} or from predominantly thermal electrons within a mildly relativistic jet \citep{FalckeMarkoff2000}. Based on observations with \textit{Chandra} \citep{Baganoffetal2003}, \citet{Quataert2002} argues that the faint quiescent X-ray emission is from thermal bremsstrahlung, originating in the outer region of the accretion disc.

While successful at producing a general description of the data, all the above semi-analytical models lack a self-consistent MHD description of the accretion flows. Although they invoke a viscosity to account for the outward angular momentum transport, in the accretion discs, they do not explicitly calculate it, nor do they account for the presence of magnetorotational instability driven accretion processes \citep{BalbusHawley1991}. GRMHD simulations are thus an ideal framework to examine the nature of accretion flows around black holes, and to test the above scenarios for Sgr A*'s emission in particular.

The GRMHD simulations presented in this work model only the innermost part of the accretion flow around Sgr A*, where the submm bump is produced. Therefore the radio (jets or outer accretion inflow) and X-ray (outer regions of the accretion disc) emission cannot be fitted by this present work and we focus our results on fitting the submm emission of Sgr A* in the quiescent state. However, we use the IR/X-ray emission as upper limits to define the feasibility of our fits.

\section{Methods}
\label{sec:methods}
The general setup of our simulations is similar to that of other groups \citep{Moscibrodzkaetal2009, Hilburnetal2010, Moscibrodzkaetal2011} in order to facilitate comparison. The simulations start with an initial torus of gas, seeded with a magnetic field, around a compact object situated at the origin. The mass of the central object is set to the mass of Sgr A* ($M_{BH} = 4.3 \times 10^{6} \mathrm{M_{\odot}}$) and the initial density profile inside the torus is chosen to produce the target mass accretion rate at the inner grid boundary. We let the simulation evolve until inflow equilibrium is established in the inner disc, where a jet, along the rotation axis, and an outward flowing wind, which we call the corona, over and under the accretion disc, are formed. See \citetalias{Dibietal2012} for more details.

To generate spectra, we had to ensure that the radiative emissivities used were physically accurate and, when integrated, produce the same cooling functions as originally included in \texttt{Cosmos++}. We thus consider the following cooling processes: bremsstrahlung, synchrotron and inverse-Compton. Since we are investigating physical processes occurring very close to a black hole, we also must include special and general relativistic effects on the radiative emission. In the following, we describe the adopted emissivity expressions.

\subsection{Radiative cooling}
\label{ssec:radiativecooling}

\texttt{Cosmos++} uses as a cooling function the following total cooling rate for an optically thin gas (\citealt{FragileMeier2009}, also see \citealt{Esinetal1996}):
\begin{align}
  \Lambda = \eta_{br,C}\, q^{-}_{br} + \eta_{s,C}\,q^{-}_s,
  \label{eqn:totalcooling}
\end{align}
where $q^{-}_{br}$ and $q^{-}_{s}$ are respectively the bremsstrahlung and synchrotron cooling terms and $\eta_{br,C}$ and $\eta_{s,C}$ are Compton enhancement factors. These $\eta$ factors are modified exponential function of the Compton parameter $y$ \citep{Esinetal1996}.

The bremsstrahlung cooling rate is taken from \citet{Esinetal1996} (equations (7) to (9) in this paper)
\begin{align}
  q^{-}_{br} = q^{-}_{ei} + q^{-}_{ee} + q^{-}_{\pm}
  \label{eqn:bremCR}
\end{align}
where $q^{-}_{ei}$, $q^{-}_{ee}$ and $q^{-}_{\pm}$ represent the cooling due respectively to electron-ion and positron-ion, electron-electron and positron-positron, and electron-positron processes.

The synchrotron cooling rate is a sum of optically thick and thin emission (equation (14) in \citet{Esinetal1996})
\begin{align}
  q^{-}_{s} = \frac{2\pi k T}{H c^2} \int^{\nu_c}_{0} \nu^2 \, \mathrm{d}\nu + \int^{\infty}_{\nu_c} \epsilon_s(\nu) \, \mathrm{d}\nu 
  \label{eqn:synchCR}
\end{align}
where $k$ is the Boltzmann constant, $T$ is the temperature of the electrons, $H$ is the local temperature scale height, $c$ is the speed of light, $\nu_c$ is the critical frequency at which the optically thick and thin emissivities are equal and $\epsilon_s(\nu)$ is the total angle-averaged synchrotron emissivity \citep{FragileMeier2009}.

The cooling rates are important to evaluate the radiative losses at each time step of the simulation.Whereas the emissivities, from which these cooling rates are derived, are the critical quantities we need to produce the spectra.

\subsubsection{Bremsstrahlung}
\label{sssec:bremsstrahlung}
Similar to \citet{Esinetal1996}, we used \citet{StepneyGuilbert1983} expression for the thermal relativistic electron-ion bremsstrahlung emissivity
\begin{align}
  \frac{dE_{ep}}{dVdtdw} = N_{p} c \int^{\infty}_{1+w} w \frac{d\sigma}{dw}\beta N_e(\gamma)d\gamma
  \label{eqn:StepneyGuilbert}
\end{align}
where $w = h\nu/m_ec^2$ is the dimensionless photon energy, $N_e(\gamma) = N_e \gamma^2 \beta \,exp(-\gamma/\theta)/\theta K_2(1/\theta)$ is the Maxwellian-J\"{u}ttner electron energy distribution, $\theta = kT/m_ec^2$ is the dimensionless temperature, and $K_2$ is a modified Bessel function. $N_e$ and $N_p$ are the electron and proton number densities, respectively.

The electron-ion cross-section needed in equation \eqref{eqn:StepneyGuilbert} can be expressed following \citet{BlumenthalGould1970}
\begin{multline}
  w \, \frac{d\sigma}{dw} = \frac{16}{3} Z^2 \alpha r_0^2 \left( 1-\frac{w}{\gamma}+\frac{3}{4} \left( \frac{w}{\gamma}\right) ^2 \right) \\  \times \left( \ln \frac{2\left( \gamma^2-\gamma w\right)}{w} -\frac{1}{2} \right)
  \label{eqn:CS}
\end{multline}
where $Z$ is the ion's atomic number, $\alpha$ is the fine structure constant and $r_0 = 2.8179 \times 10^{-13} \, \rm{cm}$ is the Compton radius.

Combining equations \eqref{eqn:StepneyGuilbert} and \eqref{eqn:CS} gives the following expression for electron-ion bremsstrahlung emissivity
\begin{multline}
  \frac{dE_{ep}}{dVdtdw} =  N_e N_p c \frac{16}{3} Z^2 \alpha r_0^2 \int^{\infty}_{1+w}\beta^2 \left( \gamma^2 - \gamma w + \frac{3}{4}w^2 \right) \\ \times \left( \ln \frac{2 \left( \gamma^2 - \gamma w \right)}{w}- \frac{1}{2}\right) \frac{e^{-\gamma/\theta}}{\theta K_2(\frac{1}{\theta})} d\gamma
  \label{eqn:bremEmis}
\end{multline}

Integrating equation \eqref{eqn:bremEmis} over frequencies leads to the bremsstrahlung cooling rate described by equation \eqref{eqn:bremCR}.

When compared to electron-ion bremsstrahlung interaction, electron-electron and electron-positron bremsstrahlung are negligible. Therefore their contribution to spectra has been ignored in the present work. Moreover, in the region of interest in our numerical simulations (i.e. $r < 15 \, r_g$, see Section \ref{ssec:geometry}), bremsstrahlung as well as Comptonization of bremsstrahlung have a smaller contribution to the overall emission in comparison with both the synchrotron and the synchrotron self-Compton processes. However, although in the case of Sgr A* bremsstrahlung emission is negligible, it will have an important contribution to the spectra of other LLAGN.

\subsubsection{Synchrotron and synchrotron self-Compton}
The total angle-averaged optically thin and thick synchrotron emissivities given by \citet{FragileMeier2009} are only valid within a certain range of temperatures. Therefore, rather than using them to account for the synchrotron contribution to the spectrum, we decided to start from first principles to express a more general expression of the synchrotron emissivity. Following \citet{RybickiLightman1986} and \citet{deKooletal1989}, we have:
\begin{align}
  I_s(\nu) = \frac{\eta_\nu}{\mu_\nu} \left( 1 - \exp(-\mu_\nu R) \right)
  \label{eqn:jnu_syn}
\end{align}
where $\eta_\nu$ is the emission coefficient, $\mu_\nu$ is the absorption coefficient and $R$ is the size of the homogeneous emitting volume.

Knowing the synchrotron radiation field $I_s(\nu)$ and following \citet{ChiabergeGhisellini1999}, the synchrotron self-Compton emissivity, in units of erg/cm$^{3}$/s/Hz/st can be expressed as
\begin{align}
  \varepsilon_c (\nu_1) = \frac{\sigma_T}{4} \int^{\nu_0^{max}}_{\nu_0^{min}} \frac{d\nu_0}{\nu_0} \int^{\gamma_2}_{\gamma_1} \frac{d\gamma}{\gamma^2 \beta^2} N(\gamma) f(\nu_0, \nu_1) \frac{\nu_1}{\nu_0} I_s(\nu_0) 
  \label{eqn:epsilon_com}
\end{align}
where $\nu_0$ is the frequency of the incident photons, $\nu_1$ is the frequency of the scattered photons, $f(\nu_0, \nu_1)$ is the spectrum produced by the single electron, scattering monochromatic photons of frequency $\nu_0$, $\beta = \frac{v}{c}$, $\nu_0^{min}$ and $\nu_0^{max}$ are the extreme frequencies of the synchrotron spectrum, and $\gamma_1$ and $\gamma_2$ are
\begin{align}
  \gamma_1 = \mathrm{max} \left[ \left( \frac{\nu_1}{4 \nu_0} \right)^{1/2}, \gamma_{min} \right]
  \label{eqn:gamma1_com}
\end{align}
\begin{align}
  \gamma_2 = \mathrm{min} \left[ \gamma_{max},\frac{3}{4} \frac{m_e c^2}{h \nu_0} \right].
  \label{eqn:gamma2_com}
\end{align}
The mean free path of inverse Compton scattering being larger than the simulation region in all our cases, we ignore multiple scatterings along the line of sight.

For the synchrotron self-Compton radiation field from a homogeneous volume of size R, equation \eqref{eqn:epsilon_com} leads to the following emissivity expression:
\begin{align}
  I_c(\nu_1) = \varepsilon_c(\nu_1) R
  \label{eqn:jnu_com}
\end{align}
In the framework of the simulation, $R$ represents the size of a zone.

Integrating, over frequencies, the synchrotron and synchrotron self-Compton radiation fields, $I_s(\nu)$ and $I_c(\nu)$, leads to the synchrotron cooling rate, with the Compton enhancement factor $\eta_{s,C}\,q^{-}_s$.

These expressions integrate to exactly the formulae used for the cooling rates within \texttt{Cosmos++}.

\subsection{General Relativistic Radiative Transfer}
The synchrotron emission and absorption coefficients (equations (4)-(5) in \citealt{deKooletal1989}), and the synchrotron self-Compton emission coefficient (equation \ref{eqn:epsilon_com}) describe the emitted spectrum from any zone in the simulation. A radiative transfer calculation is necessary to transform this into the spectrum as seen by a distant observer. Due to both strong gravitational lensing and redshifts, and Doppler beaming in the vicinity of the black hole where most of the luminosity is produced, this calculation must be done in full GR. 

The GR calculation is done using ray tracing. Starting from a distant observer's hypothetical detector, rays are traced backwards in time toward the black hole assuming they are null geodesics (geometric optics approximation), using the public code \texttt{geokerr} described in \citet{DexterAgol2009}. In the region where rays intersect the accretion flow, the radiative transfer equation is solved along the geodesic \citep{Broderick2006} in the form given in \citet{FuerstWu2004} using the code \texttt{grtrans} \citep{DexterPhD}, which then represents a pixel of the image. This procedure is repeated for many rays to produce an image, and at many observed frequencies to calculate the spectrum. 

Both gravitational redshifts and Doppler shifts lead to differences between observed and emitted frequencies. Emission and absorption coefficients are then interpolated both spatially between neighbouring zones to points on the geodesic, but also logarithmically in frequency to the emitted frequency corresponding to the desired observed frequency.

\subsection{Assumptions and numerical limitations}
\label{ssec:assumptions}
All of our models assume a thermal plasma. This plasma is described with a Maxwell-J\"{u}ttner energy distribution with temperature $T_e$, characterised by a fixed fraction of the ion temperature $T_i$. This approach is standard for MHD simulations since it would be difficult computationally to simulate two interacting plasmas. The ion temperature is calculated via the ideal gas law. Since the internal energy of the plasma is dominated by the ions, the cooling function used in the simulation is that of the ions. The assumption is made that the temperature of the electrons, $T_e$, needed in the calculation of the cooling rate (since the cooling processes that we are considering all involve electrons), is simply related to $T_i$ by a fixed factor. To get $T_e$, we assume that some process is coupling the two temperatures. In the case where the ratio is 1, we assume that the two temperatures are coupled via a perfect process. When $T_i > T_e$ this process is assumed to be imperfect. There is no reason why the plasma remains at a fixed temperature ratio throughout its evolution. However, studies have shown that allowing this ratio to be space- and time-dependent do not dramatically change the resulting simulations (based on unpublished work by Dexter). Therefore the assumption of a fixed $T_i/T_e$ may be a reasonable approximation.

We also assume that the radiation escapes freely from the system. The whole system is optically thin to synchrotron self-Compton emission while, for the calculation of the synchrotron, we consider the appropriate optical depth of the gas at a given location and time, which depends on the state of the plasma. This approximation takes into account the optical depth without performing radiative transfer, and is valid as long as the (assumed thermal) peak of the radiating particle distribution corresponds to energies greater than the self-absorption frequency, which is almost always the case for the regions under study.

Numerical caps and floors are a necessary limitation of most MHD simulations to prevent the codes from crashing in regimes where the values are too large/small. In our simulations,  floors have been applied on the matter and energy density values a zone is allowed to reach. In addition, a cap has been imposed on the magnetization of the fluid, as measured by $P_B/(\rho+\rho\epsilon)$, where $P_B$ is the magnetic pressure, $\rho$ is the density and $\epsilon$ is the internal energy. Similar floors must also be applied when post-processing the simulations. Regions of simulation that are under-dense and under-energetic (for instance inside the jets), as well as strongly magnetically dominated zones (for instance near the black hole) are not included in the spectral calculations, as they have reached these numerical floors.

\subsection{Magnetic field configuration}
\label{ssec:magnfieldconf}
Magnetorotational instabilities in MHD simulations are driven by weak poloidal magnetic field loops seeded in the initial torus. However not much is known about the magnetic field configuration in accretion discs around black holes. In the case of Sgr A*, most groups model it with one loop across the initial torus. In our work, we tested the results using two different configurations: a single set of poloidal loops (hereafter the 1-loop model) centred on the pressure maximum of the torus and following contours of pressure/density; we also run simulations with four sets of poloidal loops (hereafter the 4-loop model) spaced radially, with alternating field directions in each successive loop (see Section 5.1 of \citetalias{Dibietal2012}).

Because of the stochastic nature of MRI-generated turbulence and magnetic reconnection, MHD simulations of accretion discs can show significant variability. In addition, our axisymmetric simulations show violent flaring events triggered by reconnection. The impact of such events on the emission can be extreme, especially for a few very brief X-ray flares (see \ref{ssec:flares}).

\subsection{Spectral Energy Distribution}
\label{ssec:SED}
\begin{figure}
\centering
\includegraphics[width=84mm]{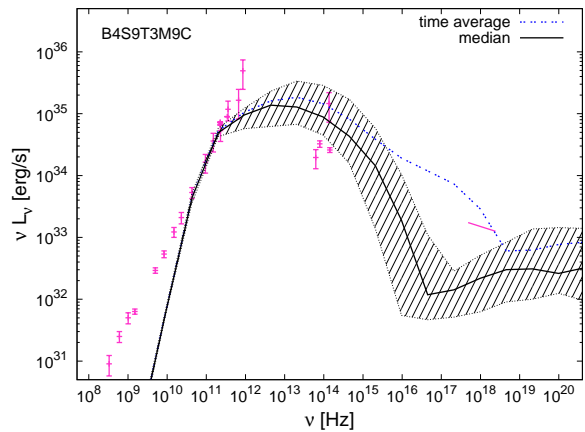}
\caption{Broadband spectra of the reference simulation B4S9T3M9C computed from all time steps in the interval 2.5-3.5 orbits. Time-averaged values (blue, dash-dash) are compared with median values (black, solid). The flaring events we see in our simulation have too much weight in the time-averaged values, which is why we have chosen to use median values to represent typical flux densities. Shadows represents the 1-sigma variability of the simulated data (see Section \ref{ssec:SED}). Observation data of Sgr A* (pink) show average quiescent spectrum published in \citet{MeliaFalcke2001}, submm data from \citet{Munozetal2012} and mean infrared from \citet{Schodeletal2011}. The X-ray is an average quiescent flux from \citet{Baganoffetal2003}.}
\label{fig:avg}
\end{figure}

We let our simulations run for 7 orbits, where we refer to the circular orbital period at $r = r_{centre}$ or $t_{orb} = 1.67 \times 10^{4}$ s. The simulations reach their targeting mass accretion rates, after their peak value and before returning to their background rates, between 2.5 and 3.5 orbits \citepalias{Dibietal2012}. To reproduce the quiescent state of Sgr A*, we take the median value of the 50 individual spectra in this interval. We do not use a time averaged SED, in order to not overweight the likely unphysical flaring episodes. Figure~\ref{fig:avg} shows that a simple time-average gives too much weight to the flaring events, increasing the flux up to an order of magnitude in the X-ray, compared to the median. The shadow region is the ``1-sigma'' variation about the median. It represents the limits within which $68 \%$ of the spectra fall. For each 50 individual spectra, the eight highest and the eight lowest data points in each spectra energy bin have been dropped.

\begin{figure}
  \centering
  \includegraphics[width=84mm]{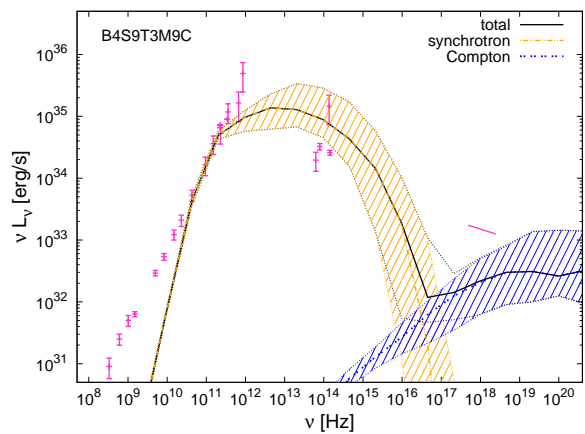}
  \caption{Broadband spectra of the B4S9T3M9C simulation presenting the synchrotron (orange, dash-dot-dot) and the synchrotron self-Compton (blue, dash-dash) components of the radiation and the resulting total emission (black, solid).}
  \label{fig:synSSC}
\end{figure}

Figure \ref{fig:synSSC} presents a sample simulated broadband spectra of Sgr A*. The first bump from the submm band to the near-infrared band is due to thermal synchrotron radiation while the second bump in the X-ray is from upscattered submm seed photons via Inverse-Compton process.

\subsection{Parameter-space}
Each model used to simulate Sgr A* is described in terms of the following five parameters: the configuration of the magnetic field $B$, the spin of the black hole $a_*$, the ion-to-electron temperature ratio $T_i/T_e$, the mass accretion rate $\dot{M}$ and enabling ($C$) or disabling the cooling function. A sixth parameter, the inclination angle $i$ at which the system is viewed from Earth is used in the ray-tracing program and has been also studied. Low inclination angle corresponds to a face-on situation while high one is edge-on. Table \ref{tab:parameters} presents our parameter-space.
\begin{table}
  \centering
  \caption{Free parameters explored in the simulations}
    \begin{tabular}{l | c}
      \hline
      Parameter			&	Values\\
      \hline
      $B$ loops ($N$)		&	1, 4\\
      Spin ($a_*$) 		&	-0.9, 0, 0.5, 0.7, 0.9, 0.98\\
      $T_i/T_e$			&	1, 3, 10\\
      Target $\dot{M}$[ M$_{\odot}$/yr]		&	$10^{-9}$, $10^{-8}$, $10^{-7}$\\
      Cooling function		&	ON ($C$), OFF\\
      \hline
      Inclination angle $i$ [deg]	&	$5^{\circ}$, $45^{\circ}$, $85^{\circ}$\\
      \hline
    \end{tabular}
    \label{tab:parameters}
\end{table}

As explained in Section~\ref{sec:obs}, there are tight observational constraints on the mass accretion rate from linear polarisation measurements in the submm band. We therefore impose boundaries of $2 \times 10^{-9} M_{\odot} \, \mathrm{yr^{-1}}$ and $2 \times 10^{-7} M_{\odot} \, \mathrm{yr^{-1}}$ to our corresponding parameter. On the other hand, despite all the data that have been gathered on Sgr A* over the years, its spin is still an unknown parameter. Therefore we have decided to explore a wide range of possible spin values, from a non-spinning black hole case ($a_* = 0$) to a maximum spinning ($a_* = 0.98$), as well as a retrograde-spinning ($a_*=-0.9$) one. Finally we allow the ion-to-electron temperature ratio vary between 1 (efficient coupling between electrons and ions) and 10 (relatively inefficient coupling), which is the same range explored in earlier works.

Throughout this paper, we follow the same naming convention defined in \citetalias{Dibietal2012} to designate simulations. Each simulation name refers to the parameter-space explored in that model. For example, B4S9T3M9C means that the initial torus is seeded with a 4-loop poloidal magnetic field, the spin of the black hole is set to $a_{*} = 0.9$, the ion-to-electron temperature ratio to 3, the mass accretion rate to $\dot{M} = 10^{-9} \mathrm{M_{\odot} \, yr^{-1}}$ and the cooling function is enabled. Table \ref{tab:sim} presents a overview of all of our simulations.

\begin{table*}
 \centering
 \begin{minipage}{140mm}
  \caption{Description of simulation parameters}
  \label{tab:sim}
  \begin{tabular}{@{}lccccccccc@{}}
  \hline
   Simulation  &  $B$ loops ($N$)    &  Spin ($a_*$)  &  $T_i/T_e$  &  Target $\dot{M}$  &  statistical av.  $\dot{M}$ &  Cooling  &  Resolution\\   
 \hline
B4S9T3M9C & 4 & 0.9 & 3 & $10^{-9}$ & 2.36$\pm 1.54 \times 10^{-9}$  & ON & $256 \times 256$ \\
B4S9T3M9Ct\footnote{Disk scale height of $H/r\approx 0.07$ instead of 0.12 as for B4S9T3M9C} & 4 & 0.9 & 3 & $10^{-9}$ & $3.05\pm2.22 \times 10^{-9}$ & ON & $256 \times 256$ \\
B4S9T3M9Cb\footnote{$\beta_\mathrm{mag,0}=50$ instead of 10} & 4 & 0.9 & 3 & $10^{-9}$ & $1.95\pm2.62 \times 10^{-9}$ & ON & $256 \times 256$ \\
B4S9T3M9Ce\footnote{$r_\mathrm{max} = 1.1\times10^4 r_G$ instead of $120 r_G$} & 4 & 0.9 & 3 & $10^{-9}$ & $3.00\pm2.00 \times 10^{-9}$ & ON & $512 \times 256$ \\
B4S9 & 4 & 0.9 & - & - & - & OFF & $256 \times 256$ \\
B1S9T3M9C & 1 & 0.9 & 3 & $10^{-9}$ & 7.93$\pm 9.27 \times 10^{-9}$ & ON & $256 \times 256$ \\
B1S9 & 1 & 0.9 & - & - & - & OFF & $256 \times 256$ \\
B4S0T3M9C & 4 & 0 & 3 & $10^{-9}$ &  2.88$ \pm 2.02 \times 10^{-9}$ & ON & $256 \times 256$ \\
B4S5T3M9C & 4 & 0.5 & 3 & $10^{-9}$ & 5.33$\pm 4.87 \times 10^{-9}$ & ON & $256 \times 256$ \\
B4S7T3M9C  & 4 & 0.7 & 3 & $10^{-9}$ & 5.38$\pm 4.06 \times 10^{-9}$ & ON & $256 \times 256$ \\
B4S98T3M9C & 4 & 0.98 & 3 & $10^{-9}$  & 3.98$\pm 3.36 \times 10^{-9}$ & ON & $256 \times 256$ \\
B4S9rT3M9C & 4 & -0.9 & 3 & $10^{-9}$ & 0.64$\pm 0.47 \times 10^{-9}$  & ON & $256 \times 256$ \\
B4S9T1M9C & 4 & 0.9 & 1 & $10^{-9}$ &  3.85$\pm 3.50 \times 10^{-9}$ & ON & $256 \times 256$ \\
B4S9T10M9C & 4 & 0.9 & 10 & $10^{-9}$ & 3.62$ \pm 3.90 \times 10^{-9}$ & ON & $256 \times 256$ \\
B4S9l & 4 & 0.9 & - & - & - & OFF & $192 \times 128$ \\
B4S9h & 4 & 0.9 & - &  - & - & OFF & $384 \times 384$ \\
B4S0T3M8C & 4 & 0 & 3 & $10^{-8}$ &  3.73 $\pm 2.18 \times 10^{-8}$ &  ON & $256 \times 256$ \\
B4S9T3M8C & 4 & 0.9 & 3 & $10^{-8}$ &  4.99 $\pm 3.80 \times 10^{-8}$ &  ON & $256 \times 256$ \\
B4S9T3M7C  & 4 & 0.9 & 3 & $6.3\times 10^{-8}$ & 1.69 $\pm 1.39 \times 10^{-7}$ & ON & $256 \times 256$ \\
B4S0T3M7C & 4 & 0 & 3 & $6.3\times 10^{-8}$ & 2.16 $\pm 1.46 \times 10^{-7}$ & ON & $256 \times 256$ \\
B4S5T3M7C  & 4 & 0.5 & 3 & $6.3\times 10^{-8}$ & 2.03$ \pm 1.85 \times 10^{-7}$  & ON & $256 \times 256$ \\
B4S75T3M7C  & 4 & 0.75 & 3 & $6.3\times 10^{-8}$ & 1.65$ \pm 1.15 \times 10^{-7}$ & ON & $256 \times 256$ \\
B4S98T3M7C  & 4 & 0.98 & 3 & $6.3\times 10^{-8}$ & 1.37$\pm 1.08 \times 10^{-7}$  & ON & $256 \times 256$ \\
B4S9rT3M7C  & 4 & -0.9 & 3 & $6.3\times 10^{-8}$ &  0.39$\pm 0.41 \times 10^{-7}$ & ON & $256 \times 256$ \\
B4S9T1M7C  & 4 & 0.9 & 1 & $6.3\times 10^{-8}$ & 1.53$\pm 1.02 \times 10^{-7}$ & ON & $256 \times 256$ \\
B4S9T10M7C  & 4 & 0.9 & 10 & $6.3\times 10^{-8}$ & 1.43$ \pm 0.88 \times 10^{-7}$ & ON & $256 \times 256$ \\
\hline
\end{tabular}
\end{minipage}
\end{table*}

B4S9T3M9C at an inclination angle of $85^{\circ}$ is the closest solution to those found by previous works when attempting to fit Sgr A*'s data with simulated SEDs, except for the enabling of the cooling function, and the initial magnetic field configuration. In the following we have chosen this set of initial parameters to be our reference simulation, from which we have explored our parameter space.

\section{Results}
\label{sec:results}
\subsection{Geometry}
\label{ssec:geometry}
\begin{figure}
  \centering
  \resizebox{60mm}{!}{\includegraphics[width=84mm]{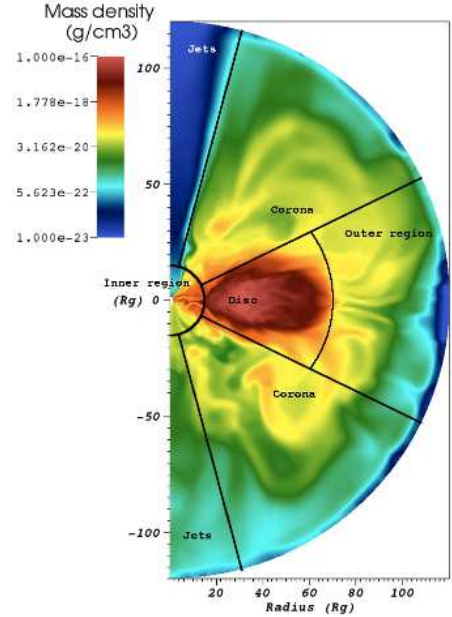} }
  \caption{Delimitation of the different regions of our simulations considered in the simulated spectra. The limits of each region are: inner region [event horizon $< r < 15 \, r_g$] - disc [$15 \, r_g < r < 70 \, r_g$ : $0.36\pi < \theta < (1-0.36)\pi$] - outer region [$70 \, r_g < r < 120 \, r_g$ : $0.36\pi < \theta < (1-0.36)\pi$] - corona [$15 \, r_g < r < 120 \, r_g$ : $0.08\pi < \theta < 0.36\pi$] - jets [$15 \, r_g < r < 120 \, r_g$ : $0.004\pi < \theta < 0.08\pi$]. The corona and jet regions are symmetric with respect to $\theta = \pi/2$.}
  \label{fig:geoGRMHD}
\end{figure}

\begin{figure}
  \centering
  \includegraphics[width=84mm]{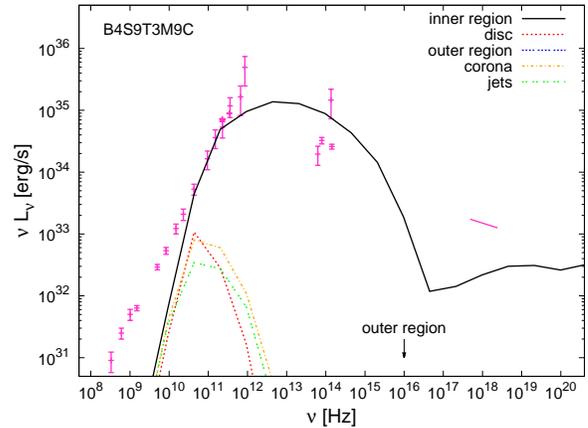}
  \caption{Broadband spectra comparing the emission from every part of the simulation as shown in Figure~\ref{fig:geoGRMHD}. For clarity, the 1-sigma variability has been omitted.}
  \label{fig:geoSED}
\end{figure}

To compute our spectra, we had to ensure that only regions of the simulations that have reached inflow equilibrium are contributing to the emission. The reason is that properties of the parts of the simulations that have not reached the inflow equilibrium are strongly dependent on the arbitrary, initial conditions. As a consequence, only radiation from a region lying between the event horizon and 15 $r_g$ are accounted for in our spectra \citepalias{Dibietal2012}. Fig.\ref{fig:geoGRMHD} illustrates this selection. This geometrical restriction is consistent with our aim to only fit the submm bump, which is believed to originate very close to the black hole. Indeed, Figure~\ref{fig:geoSED} shows that this inner region of the accretion flow accounts for the bulk of radiation in our simulation.

\subsection{Exploring the parameter space}
\begin{figure}
  \centering
    \includegraphics[width=84mm]{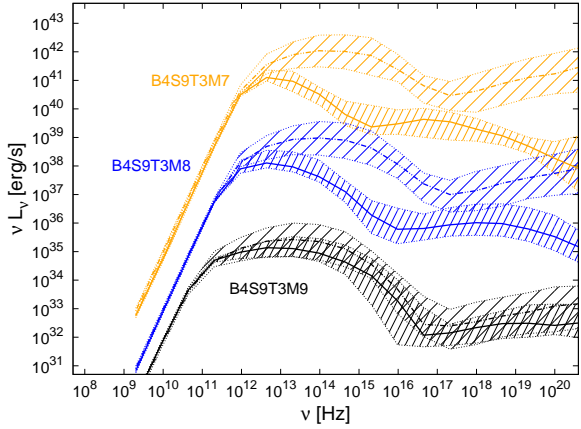}
    \caption{Broadband spectra comparing, at three different mass accretion rates: $10^{-9}$ M$_{\odot}$/yr (black), $10^{-8}$ M$_{\odot}$/yr (blue), and $10^{-7}$ M$_{\odot}$/yr (orange), the effect of turning ON (solid) and OFF (dash-dot-dot) the cooling function in simulations with the same set of initial parameters (4-loop model, $a_* = 0.9$, $T_i/T_e = 3$ and $i=85^{\circ}$). Emission from simulations B4S9T3M8 and B4S9T3M7 have been respectively multiplied by 10 and 1000 to improve readability.}
    \label{fig:Cooling}
\end{figure}

For the first time, we are able to generate consistent spectra from GRMHD simulations that can be compared to observations in a robust way, as no post-processing scaling is possible when the cooling function is enabled. To assess the importance of self-consistent treatment of radiative losses on the resulting emission, we have compared SEDs from simulations with the same set of initial parameters (4-loop model, $a_* = 0.9$ and $T_i/T_e = 3$), with and without radiative cooling. This comparison was done for three different target mass accretion rates. Figure~\ref{fig:Cooling} presents the six spectra obtained. When enabling the cooling function, there is a clear trend of increasing importance of the effect with increasing mass accretion rate. While at a mass accretion rate of $10^{-9}$ M$_{\odot}$/yr, both spectra of cooling and non-cooling simulations are similar, significant differences (up to two orders of magnitudes) appear at a mass accretion rate of $10^{-8}$ M$_{\odot}$/yr and $\dot{M} = 10^{-7}$ M$_{\odot}$/yr.

\begin{figure}
  \centering
\includegraphics[width=84mm]{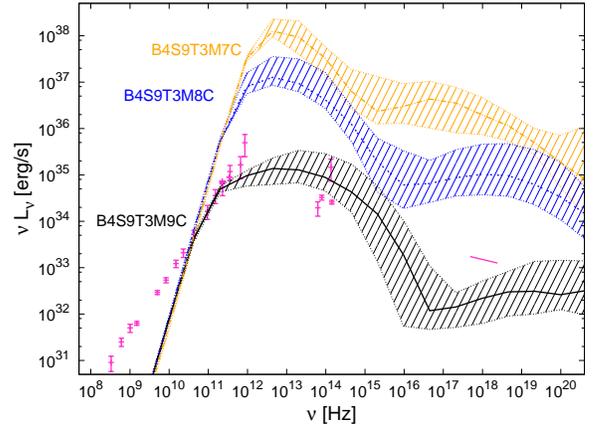}
\caption{Broadband spectra comparing the effect on emission of varying the mass accretion rate, when the cooling function is turned on . All simulations have been set with an initial 4-loop magnetic field, a black hole spin $a_* = 0.9$, a temperature ratio $T_{i}/T_{e} = 3$ and an inclination angle $i=85^{\circ}$.}
\label{fig:Mdot}
\end{figure}

Next, we have naturally focused our attention on how varying the mass accretion rate changes the resulting spectra. As shown in Figure~\ref{fig:Mdot}, which compares SEDs from simulations B4S9T3M9C, B4S9T3M8C and B4S9T3M7C, there is a significant positive correlation between the mass accretion rate and the emission at all wavelengths.

\begin{figure}
  \centering
\includegraphics[width=84mm]{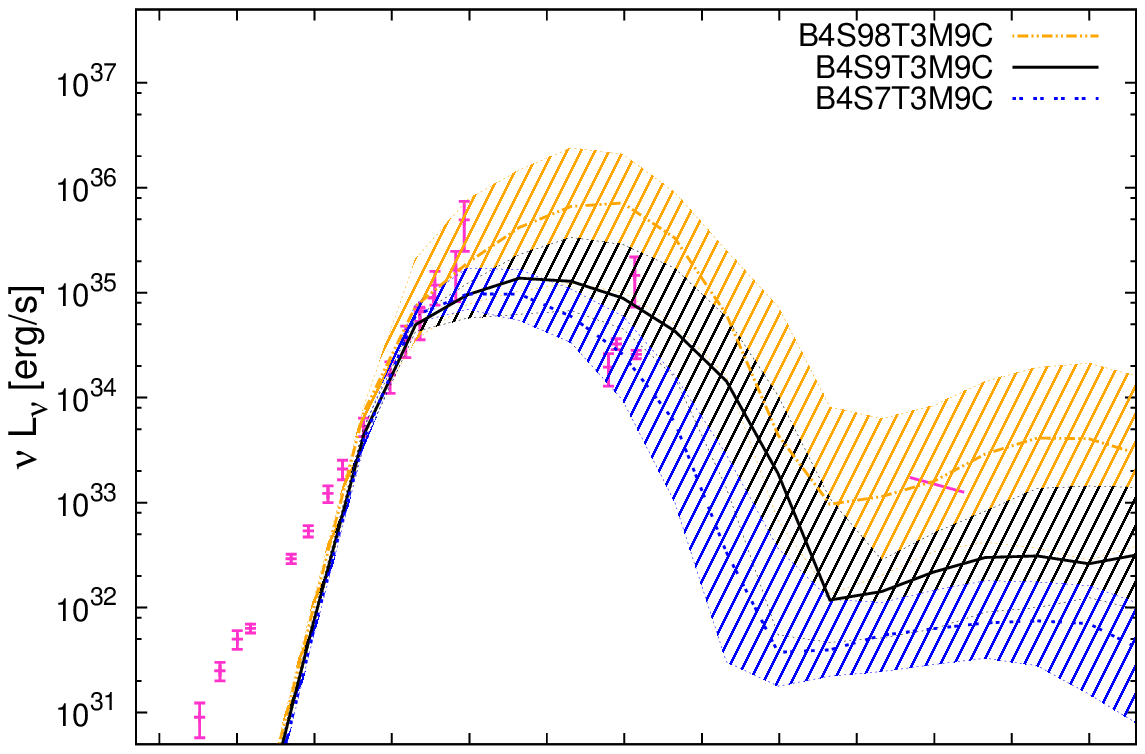}
\includegraphics[width=84mm]{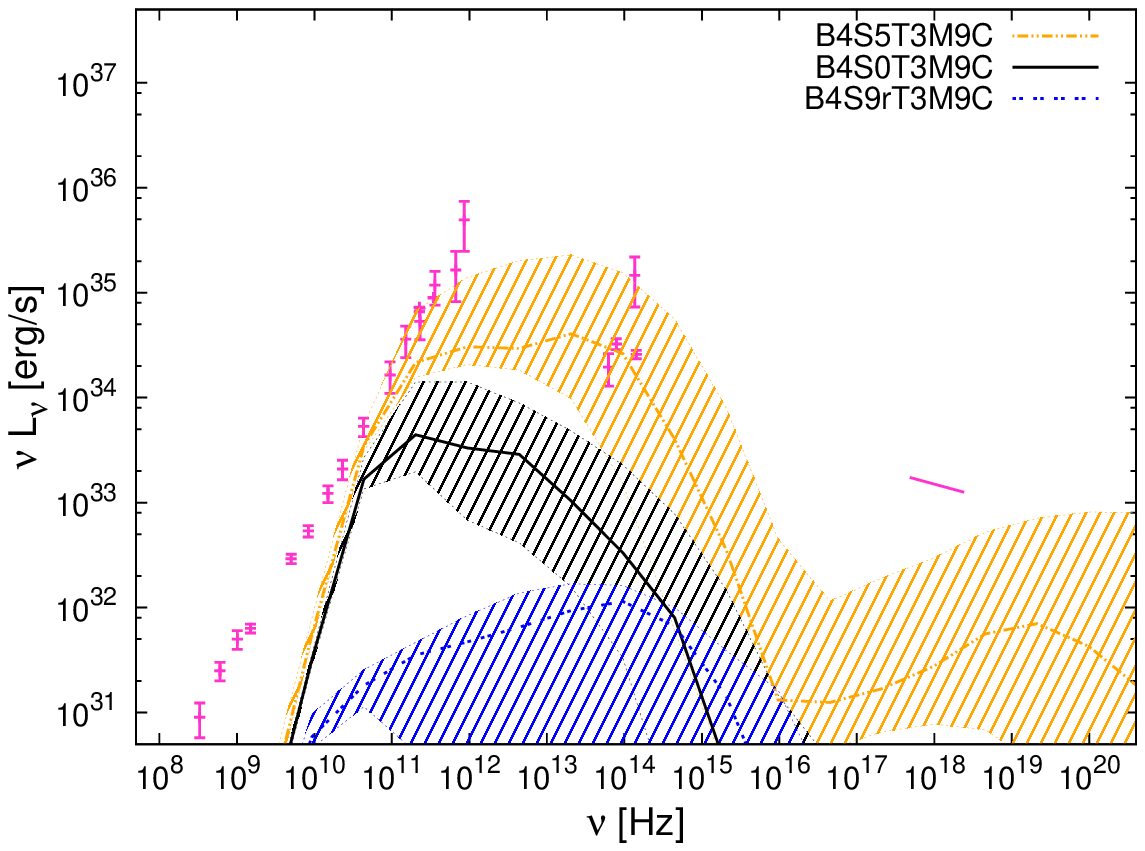}
\caption{Broadband spectra of simulations with the same set of initial parameters (4-loop model, $T_i/T_e = 3$, mass accretion rate $\dot{M} = 10^{-9}$ M$_{\odot}$/yr and $i=85^{\circ}$) but for different black hole spin values: $a_* = \{0.7,0.9,0.98\}$ (top panel), $a_* = \{-0.9,0,0.5\}$ (bottom panel).}
\label{fig:spin}
\end{figure}

We also test the role of black hole spin $a_*$. As can be seen from Figure~\ref{fig:spin}, the luminosity rises from the case of a retrograde spinning black hole to a spinning one of $a_* =$ 0.5. Then, it appears that the emission reaches a plateau between $a_* =$ 0.5 and $a_* =$ 0.7, before rising again. Moreover, simulations from models with spin $a_* =$ -0.9, 0, 0.5 and 0.98 seem to be more variable compared to models of spin $a_* =$ 0.7 and 0.9, which lead to a higher variability in the resulting radiation. Finally comparing emission from positive and negative spins reveals almost four orders of magnitude difference between fluxes at same absolute spin value.

\begin{figure}
  \centering
\includegraphics[width=84mm]{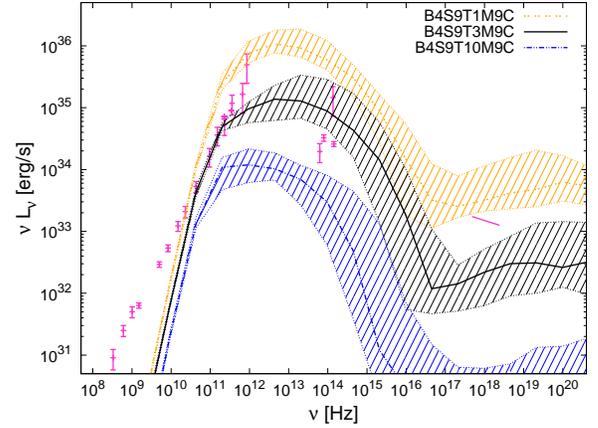}
\caption{Broadband spectra comparing the effect of varying the ion-to-electron temperature ratio for simulations with an initial 4-loop magnetic field, a black hole spin $a_* = 0.9$, a mass accretion rate $\dot{M} = 10^{-9} M_{\odot} \mathrm{yr^{-1}}$ and an inclination angle $i=85^{\circ}$.}
\label{fig:temp_effect}
\end{figure}

Regarding the ion-to-electron temperature ratio, it is straightforward to assess how this parameter affects the radiation in non-cooling simulations, but it is not for cooling ones. On one hand, in non-cooling simulations, the temperature of the ions stays the same. So increasing the temperature ratio decreases the temperature of the electrons and therefore the emission. On the other hand, in cooling simulations, effects from radiative losses and from efficiency of cooling processes between ions and electrons conflict and their results on spectra are not straightforward. Radiative losses will lower the temperature of the electrons while increasing $T_i/T_e$ results in having less efficient cooling processes, therefore less radiative emission from the electrons and thus a higher electron's temperature. Nonetheless Figure~\ref{fig:temp_effect} shows that in our specific case, increasing $T_i/T_e$ decreases the total emission. However, at higher accretion rates this increase is less than the $F_\nu \sim \dot{M}^2$ scaling without cooling.

\begin{figure}
  \centering
\includegraphics[width=84mm]{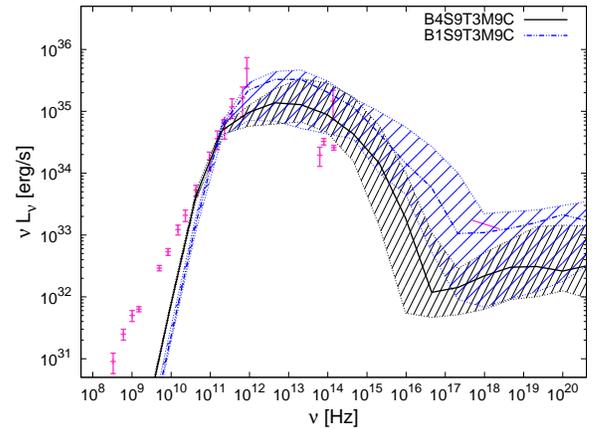}
\caption{Broadband spectra comparing the effect of varying the initial magnetic field configuration on the emission of simulations with a black hole spin $a_* = 0.9$, a temperature ratio $T_{i}/T_{e} = 3$, a mass accretion rate $\dot{M} = 10^{-9} M_{\odot} \mathrm{yr^{-1}}$ and an inclination angle $i=85^{\circ}$.}
\label{fig:Bconfig}
\end{figure}

Our work is one of the first to address the question of how magnetic field configuration model in the initial accretion disc affects the resulting emission, in the case of Sgr A*. Figure~\ref{fig:Bconfig} compares spectra from two magnetic field configurations: the 1-loop and the 4-loop models. The figure shows that, while in the submm and the near-infrared bands the emission is fairly independent of the model -- both spectra are within each others variability range -- in the X-ray, the emission is very sensitive to the initial configuration.

\begin{figure}
  \centering
\includegraphics[width=84mm]{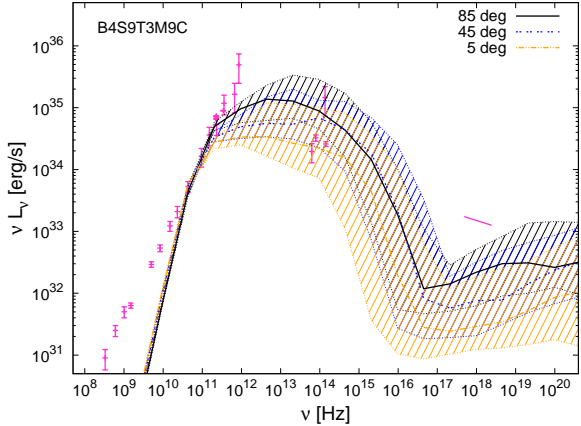}
\caption{Broadband spectra comparing the effect of varying the viewing inclination angle on the emission of the reference simulation B4S9T3M9C.}
\label{fig:inclination}
\end{figure}

The inclination angle is the last parameter we tested. This parameter is only used in the post-processing ray-tracing code \texttt{grtrans}. It defines the viewing angle of a distant observer on the system. Doppler shifts and optical depth are the two significant changes induced by varying the inclination angle \citep[e.g.,][]{Dexteretal2009}. For most observer inclinations, Doppler beaming is the predominant effect from ray-tracing from Keplerian discs while at lower inclinations optical depth is the main effect. Figure~\ref{fig:inclination} shows that, at higher inclinations, Doppler beaming leads to larger fluxes, and moves the peak of the spectrum to higher frequency. At lower inclinations, the figure also shows that optical depth causes larger variability of the overall fluxes.

\section{Discussion}
\label{sec:discussion}
\subsection{Preferred parameter space}
\begin{figure}
  \centering
\includegraphics[width=84mm]{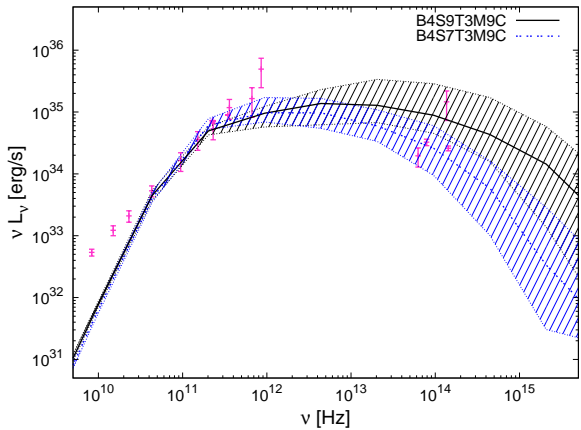}
\caption{Broadband spectra of our preferred parameters fits. The models suggest that the mass accretion rate at which Sgr A* accretes is around $10^{-9} M_{\odot}$/yr and the spin of the central black hole is likely to be between $a_* = 0.7$ and $a_* = 0.9$.}
\label{fig:submm_fit}
\end{figure}
The present study was designed to determine which set of parameters will be the closest to reproducing the quiescent state of Sgr A*. Our results show that the most compatible spectra with the observational data are those of simulations B4S9T3M9C and B4S7T3M9C. We find that we can fit the Sgr A* data at 230 GHz, which has a value of 3 Jy, or $\nu L_{\nu} \sim 5.6 \times 10^{34}$ erg/s, with a mass accretion rate of $2.60 \pm 1.54 \times 10^{-9} M_{\odot}$/yr in simulation B4S9T3M9C and $5.38 \pm 4.06 \times 10^{-9} M_{\odot}$/yr in simulation B4S7T3M9C. These values confirm the conclusion of \citetalias{Dibietal2012} that it is not necessary to consider radiative processes when simulating accretion onto Sgr A* because of its exceptionally low accretion rate.

Our favored target mass accretion rate is consistent with the lower limit imposed by observations of linear polarisation in the submm bump. It is interesting to note that, at this mass accretion rate, two models provide a good description of the data, with their only difference being the spin parameter value. Figure~\ref{fig:submm_fit} shows that our models with $a_* = 0.7$ and $a_* = 0.9$ match Sgr A* data at 230 GHz and  $\sim 5.6 \times 10^{34}$ erg/s. These two models distinguish themselves only in the description of the near-infrared observations. Our results are consistent with the upper limit of $a_* = 0.86$ at $2 \sigma$ significance given by \citet{Brodericketal2011} obtained with millimetre-VLBI observations \citep{Doelemanetal2008,Fishetal2011}.

As reported in \citet{Dexteretal2010}, the favored ion-to-electron temperature ratio depends strongly on initial conditions since the temperature of the ions scales with the disc thickness. The real constraint is therefore on the temperature of the electrons. Currently, our ability to constrain this parameter is only as good as the code itself. Although our preferred parameter-space fits are found with a ion-to-electron temperature ratio of 3 which suggests that the processes coupling the ions to the electrons in the accretion disc are mildly inefficient, we would advise caution regarding this conclusion.  Similarly, no strong constraints can be drawn from our study of the inclination angles. Our preferred parameters fit is obtained for an inclination angle of $ 85 \pm 40 \, \mathrm{deg}$, which is not unexpected given our position in the plane of the Galaxy.

A further interesting point is that the X-ray emission is very sensitive to the initial magnetic field configuration in our simulations, while the submm emission is fairly independent. The X-ray upper limit of Sgr A*'s data may be a promising way to constrain the effect of magnetic fields. We only tested a limited set of initial conditions, and in this context, we obtained our best fit when seeding the initial torus with a four sets of poloidal magnetic field loops.

\subsection{Cases of retrograde spin}
\label{ssec:spin}
\begin{figure}
  \centering
  \includegraphics[width=84mm]{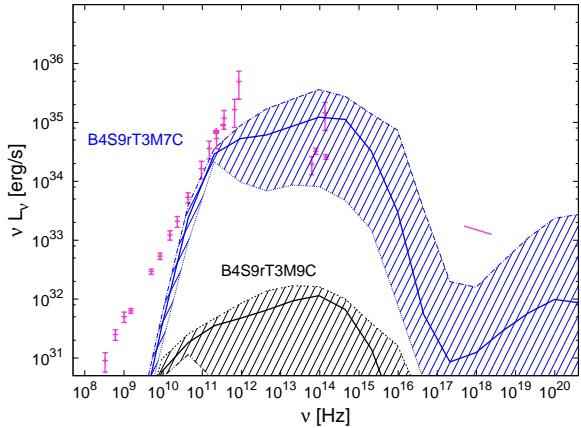}
\caption{Broadband spectra of retrograde spin simulations with the same set of initial parameters (4-loop model, $T_i/T_e = 3$ and $i=85^{\circ}$) but at different mass accretion rates.}
  \label{fig:spinR}
\end{figure}

Figure~\ref{fig:spinR} shows the spectra of retrograde spin models at mass accretion rates of $10^{-9} M_{\odot} \, \mathrm{yr^{-1}}$ and $10^{-7} M_{\odot} \, \mathrm{yr^{-1}}$. It is interesting to note that the spectral shapes of these models are significantly different from the positive spin models. The most notable difference is that the synchrotron emission peaks at higher frequency. This peak originates from an emitting region inside $3 \, \mathrm{r_g}$, filled with material at a very high temperature ($\sim 10^{13} \, \mathrm{K}$). While this very high temperature and the displacement of the synchrotron peak might be seen as criteria to exclude $a_* < 0$ models for Sgr A*, based on the submm VLBI data, it is likely premature to draw any definite conclusions. Further analysis is required to understand the origin of the very hot material and to establish whether a retrograde spin model for Sgr A* with a high mass accretion rate is viable.

\subsection{Flaring events}
\label{ssec:flares}
\begin{figure}
  \centering
  \includegraphics[width=84mm]{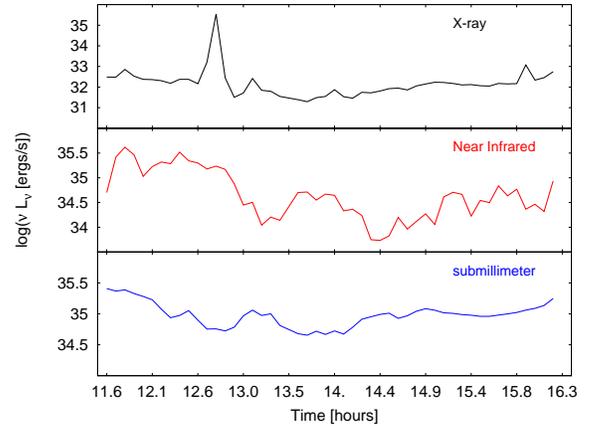}
  \caption{Light curves at three different bands: $2.1\times10^{17} \, \mathrm{Hz}$ (top), $4.5\times10^{14} \, \mathrm{Hz}$ (middle) and $9.4\times10^{11} \, \mathrm{Hz}$ (bottom) of the simulation B4S9T3M9C.}
  \label{fig:lightCurves}
\end{figure}

\begin{figure}
  \centering
  \includegraphics[width=84mm]{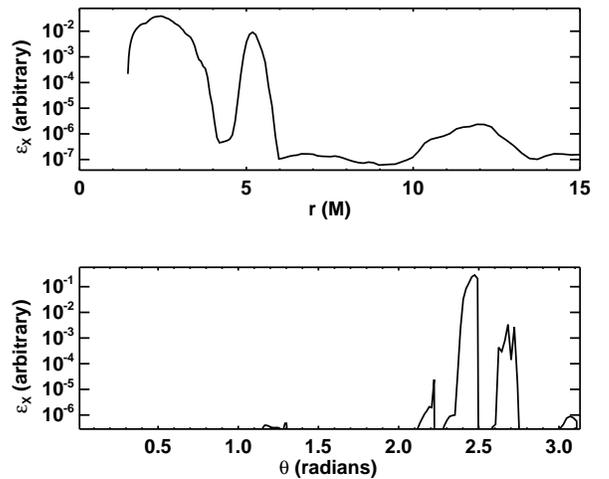}
  \caption{Radial and $\theta$ profile of the logarithm of the X-ray emissivity, area and redshift weighted. The emissivity is calculated at the peak of the X-ray flares shown in the top panel of Figure~\ref{fig:lightCurves}.}
  \label{fig:profiles}
\end{figure}

Although the main goal of our work was to reproduce the quiescent state of Sgr A*, we want to say a few words about the origin of flaring events leading to fast variability within our spectra.  As shown in Figure~\ref{fig:lightCurves}, our preferred simulation (B4S9T3M9C) experiences a brief X-ray flare, with the emission in this band increasing by three orders of magnitude over $\sim 20$ minutes. It is apparent from Figure~\ref{fig:profiles}, which presents the radial and the $\theta$ angle profile of the peak of the X-ray emission, that this emission originates from a narrow region located between $2.3 \, \mathrm{rad} < \theta < 2.8 \, \mathrm{rad}$, composed of two blobs: the first  situated around $r = 5 \, \mathrm{r_g}$ and the second  spreading between $10 \, \mathrm{r_g}$ and $14 \, \mathrm{r_g}$.

\begin{figure}
  \centering
  \includegraphics[width=40mm]{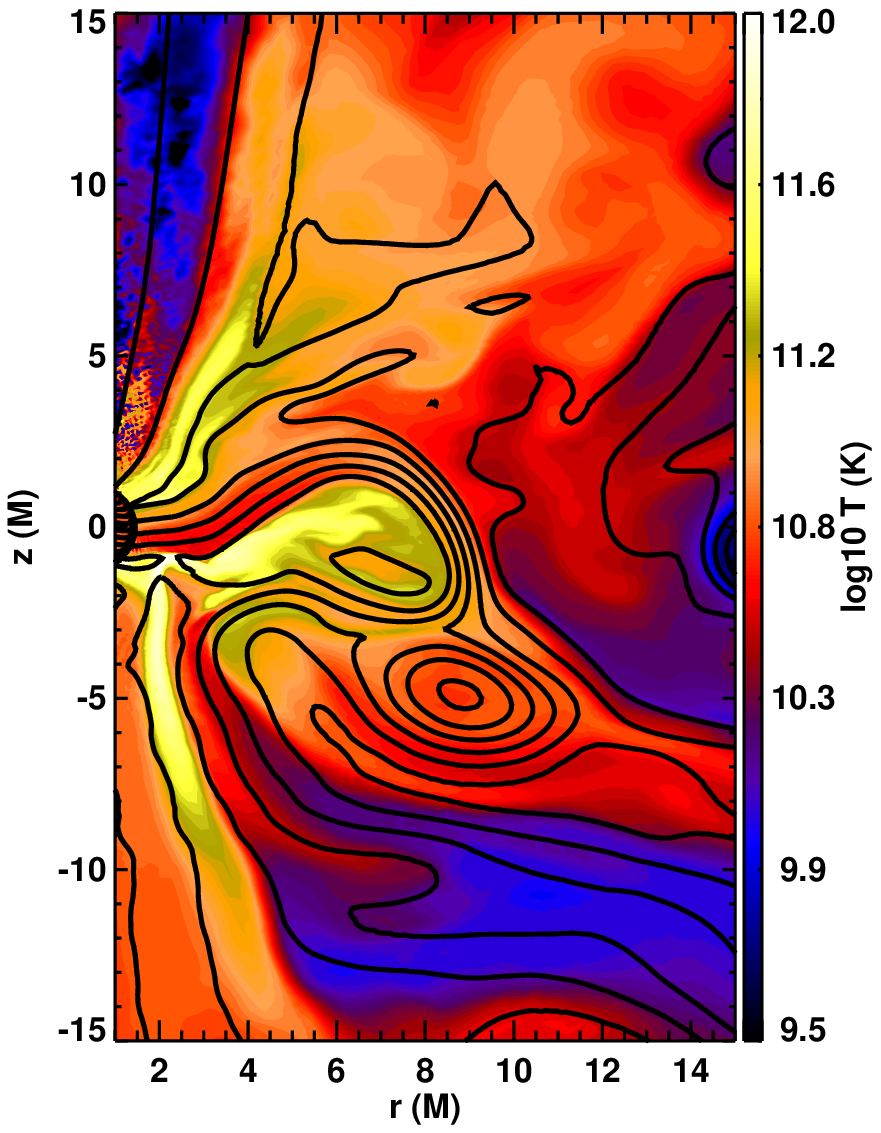}
  \includegraphics[width=40mm]{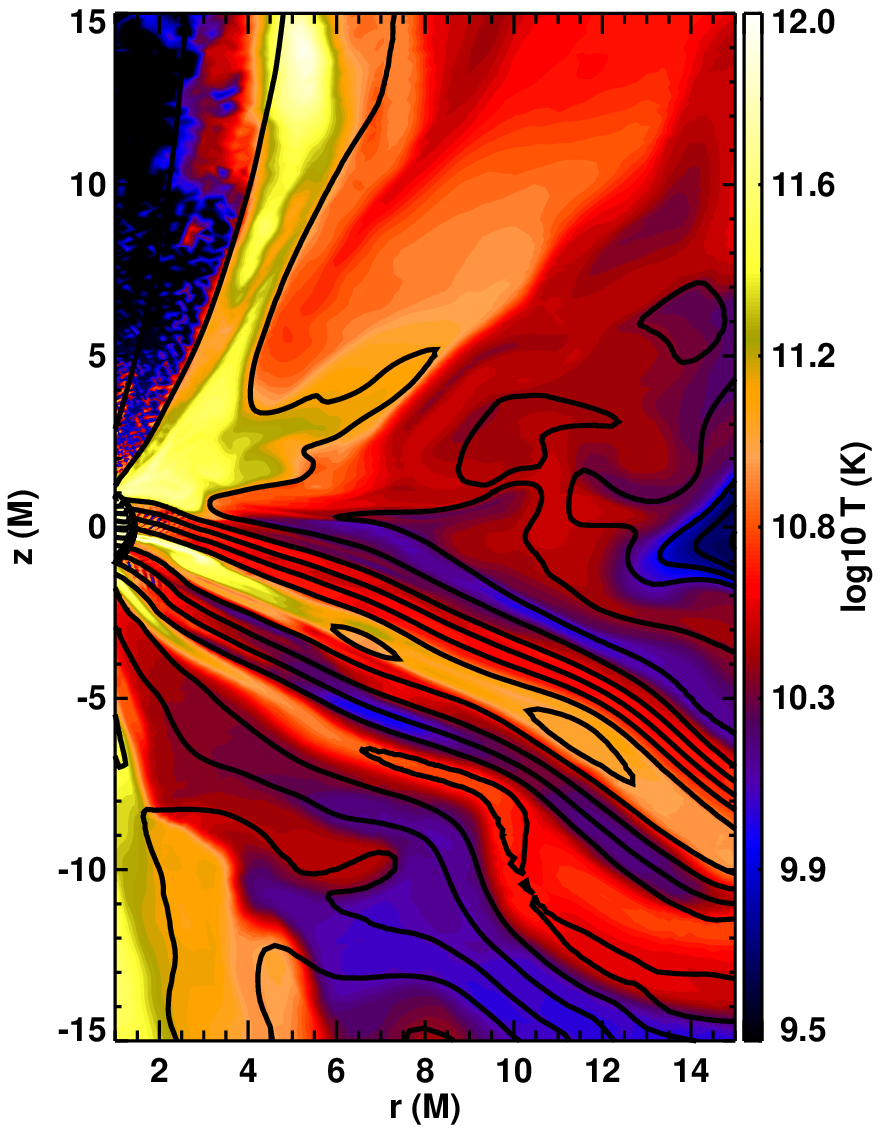}
  \includegraphics[width=40mm]{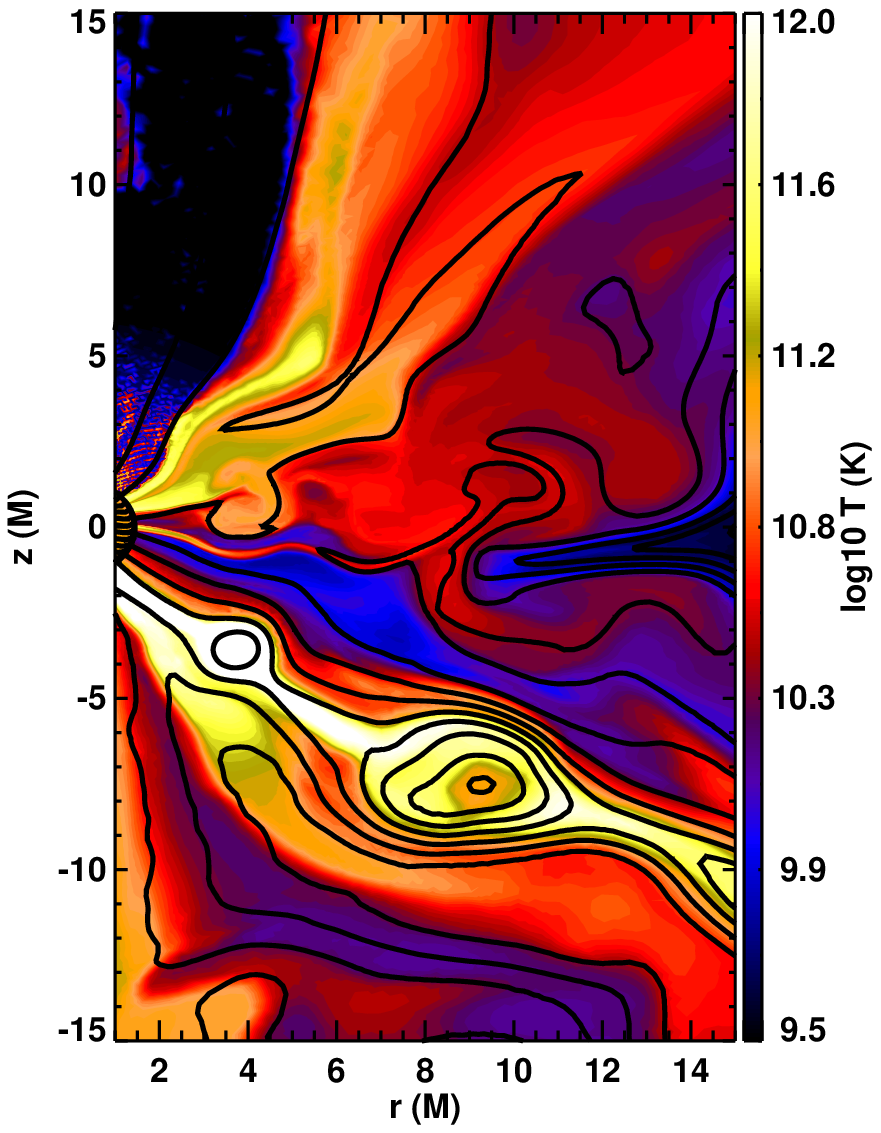}
  \includegraphics[width=40mm]{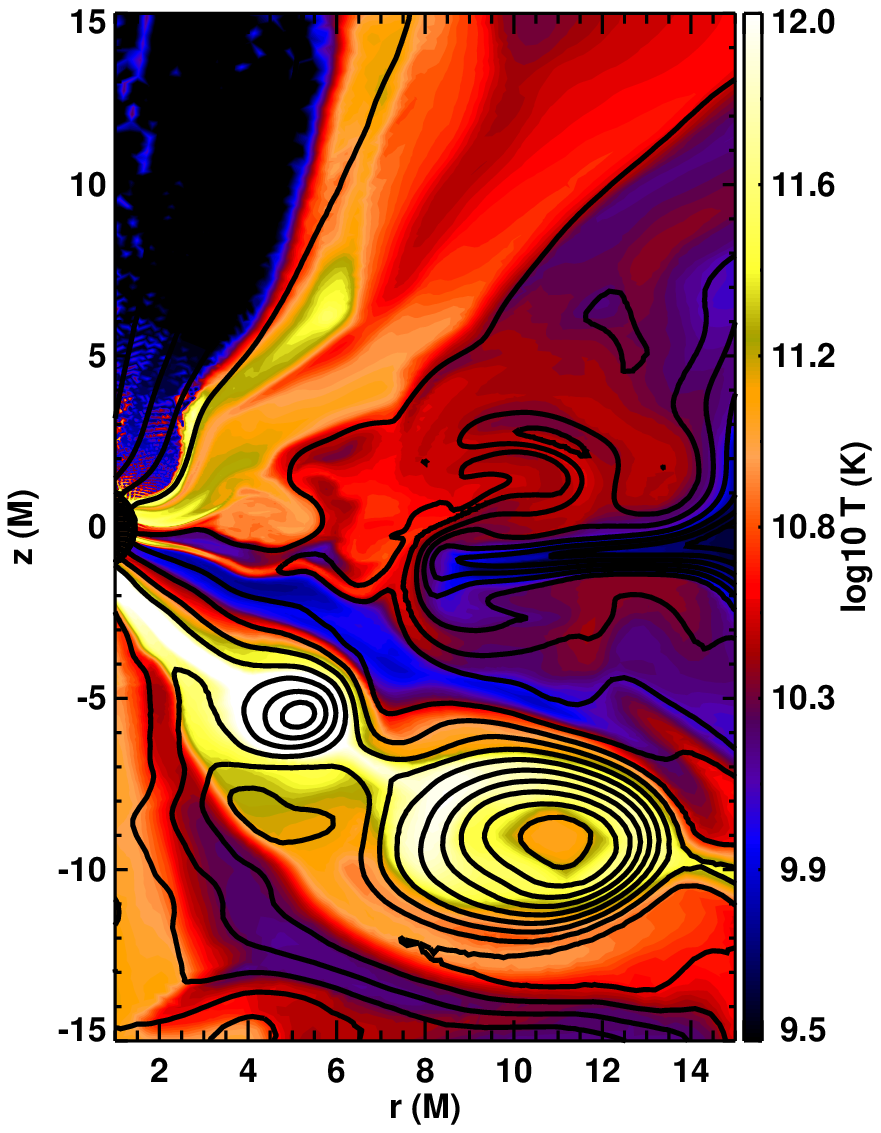}
  \caption{Snapshots of a flaring event occurring in simulation B4S9T3M9C at 11.73h (top-left), 12.381h (top-right), 12.754h (bottom-left) and 12.847h (bottom-right). Each snapshot shows a map of the temperature (color) and magnetic field lines (black lines).}
  \label{fig:aphiTemperature}
\end{figure}

To investigate the origin of the blobs, we examined temperature maps at each time step of the simulation together with the evolution of its magnetic field lines. Figure~\ref{fig:aphiTemperature} presents four snapshots of the formation of the blobs. The initial state of the flare region is a thin filament of coherent field starting at the event horizon and extending out to $r \sim 15 \, \mathrm{r_g}$, which suggests that the simulation develops a channel-mode solution \citep{HawleyBalbus1992} in this region.

These episodic flares in our 2.5D simulations are very similar to what \citet{DoddsEdenetal2010} report in their study of large, sporadic magnetic reconnection events occurring near the last stable circular orbits in their 2.5D GRMHD simulations. They suggest that because these events have timescales and energetics consistent with Sgr A*'s flares, they may represent actual physical mechanisms.  However, Sgr A*'s X-ray flares always have a simultaneous infrared counter-part, while only the largest infrared flares show X-ray flares in general \citep{Eckartetal2006,DoddsEdenetal2011}. Our simulated light curves do not show an infrared event corresponding to the X-ray flare. Moreover, no similar behaviour has ever been reported in 3D GRMHD simulations thus it is highly likely that these flaring events are numerical artefacts rising from the two dimensional nature of the simulations when magnetic reconnections occur near the event horizon of the black hole. By enforcing axisymmetry, 2.5D simulations allow larger coherent magnetic field structures to form, enhancing variability when these structures finally reconnect. We choose to minimize the effect of these rare events on the final spectra by using median rather than time-averaged spectra as discussed above.

\subsection{Comparison with previous works}
In the past few years, several groups have focused their analysis on comparing simulated observation of Sgr A* to data. While the general setup of our simulations are similar to that of other groups, the details of the code itself and the treatment of radiation are not. Somewhat surprisingly, and encouragingly, this study produces results which corroborate the findings of a great deal of the previous work in this field, suggesting that all groups are converging on a consistent picture for conditions around the supermassive black hole.  In particular, our work is in agreement with the findings of \citet{Moscibrodzkaetal2009} which showed that a radiative model of Sgr A* with a mass accretion rate of $1.86 \times 10^{-9} M_{\odot} \, \mathrm{yr^{-1}}$, an ion-to-electron temperature ratio of 3, a spin of $a_* =$ 0.94 and an inclination angle of $85 \pm 40 \, \mathrm{deg}$ fits Sgr A* observations.

These findings are also consistent with those of \citet{Dexteretal2009,Dexteretal2010} and further support the idea that the submillimeter bump in Sgr A*'s data originates from within the innermost region of an accretion disc, accreting on Sgr A* at a mass accretion rate of $\sim 2 \times 10^{-9} M_{\odot} \, \mathrm{yr^{-1}}$. However, it is worth noting that all simulations so far are finding the submm bump to be dominated by the emission from the inner disc because the jets are not yet correctly physically described.  Most importantly, idealized MHD prevents realistic mass loading in the jet funnels, but the resolution of the grid is often poor along the poles, and the numerical floors often dominate in these regions of the simulations as well.

Another important result from this work is a quantitative measure of the increasing importance of a self-consistent treatment for radiative cooling losses in GRMHD simulations, with increasing mass accretion rates. This result supports our companion paper \citetalias{Dibietal2012}'s conclusion that above a mass accretion rate of $\sim 10^{-7} \dot{M}_{\mathrm{Edd}}$, a self-consistent treatment of the radiative losses in GRMHD simulations not only affects the dynamics of the simulations, it also affects the radiative emission. This conclusion will affect any previous works done on Sgr A* which neglected the radiative losses and used a mass accretion rate higher than this limit. For example, \citet{Shcherbakovetal2010} fits Sgr A* data in the submillimeter bump and uses polarised radiation to find a mass accretion rate of $(1.4-7.0) \times 10^{-8} M_{\odot} \mathrm{yr^{-1}}$. Our result here implies that their final spectrum would be affected by cooling losses.

\subsection{Limitations}
The current study has significant limitations. All of the GRMHD simulations presented here are axisymmetric (2.5D). Axisymmetric simulations cannot sustain turbulence and so never reach a quasi-steady state. Axisymmetry also tends to exaggerate variability relative to the 3D case, and is likely responsible for the rare, large amplitude flaring events seen in many of our simulations.

Another limitation of our study, shared in general by the current class of ideal MHD simulations, is that the jets cannot be mass-loaded. Observations of flat/inverted spectra from compact jets in LLAGN, indicate optical depth effects which the current simulations cannot approach. Most likely once prescriptions for mass loading and particle acceleration in the jets are included, emission from the base of the jets will increase in the submm for Sgr A*, and have some effect on our favored parameter space.

We have also simulated a limited set of initial conditions. We have found that the initial magnetic field configuration can have an important effect on the resulting spectra, especially at high energies, but a wider range of configurations should be tried to fully explore this issue. For instance, \citet{McKinneyetal2012} argue that the initial condition used in these simulations artificially restricts the available magnetic flux, and show that large amounts of coherent flux can significantly alter the dynamics of the accretion flow.

Finally, this and almost all previous studies attempting to constrain the parameters of Sgr A* have assumed that the accretion flow angular momentum axis is aligned with the black hole spin axis. However, this is unlikely to be the case in reality, and \citet{DexterFragile2012} show that spectral fits can change dramatically even for tilts as small as $15^\circ$.

\section{Summary}
\label{sec:summary}
This paper presents for the first time self-consistent spectra from radiatively cooled GRMHD simulations of the accretion flow around a black hole, in particular, Sgr A*.  Our study concludes that the central black hole is most likely rapidly spinning ($0.7 < a_* < 0.9$), and that Sgr A* is accreting at a mass accretion rate of $\sim 2 \times 10^{-9} M_{\odot} \, \mathrm{yr^{-1}}$. While no significant conclusions can be drawn from the correlation between the resulting emission and the initial magnetic field configuration model, we obtain our best description for the submillimeter data by seeding the initial torus with a 4-loop poloidal magnetic field, suggesting that a more complex morphology could be favoured. Finally no constraints on the inclination angle can be derived from our work, but it is consistent with the general sense that Sgr A* should be more edge on than face on.  

Our work confirms the limit on the mass accretion rate ($\sim 10^{-7} \dot{M}_{Edd}$) reported in our companion paper \citetalias{Dibietal2012}, where self-consistent treatment of cooling losses in GRMHD simulations becomes important. Above this limit, spectra generated from GRMHD simulations where radiative losses are not taken into account can be potentially orders of magnitude too high. However, for other sources the exact limit may vary slightly with the mass and spin as well as initial conditions of the simulation. Nonetheless, this result is very important to keep in mind for future studies of more typical nearby LLAGN such as M81, M87, etc..

We showed that high energy emission from GRMHD simulations is sensitive to the magnetic field configuration in the initial accretion disc. Further research regarding the role of the magnetic field configuration in the dynamics and radiation of GRMHD simulations may ultimately help distinguish between models for the origin of the magnetic fields close to the black hole.

Recently there have been claims of a 3 pc-scale, jet-driven outflow from Sgr A* in the radio \citep{YusefZadehetal2012} as well a large-scale jet feature in the \textit{Fermi} GeV $\gamma$-ray maps of the Galactic center \citep{SuFinkbeiner2012}.  If one or both can be confirmed, these features will provide valuable constraints for the next technological development of GRMHD simulations, which is the inclusion of more realistic mass-loading and particle acceleration in the jets. Similarly, the discovery of the G2 cloud \citep{Gillessenetal2012} on a collision course with Sgr A* for 2013 may provide new tests of Sgr A*'s emission at higher accretion rates, for comparison with our results here.

\section*{Acknowledgements}
S.Dr. and S.M. acknowledge support from a Netherlands Organisation for Scientific Research (NWO) Vidi Fellowship. S.Di. and S.M. also acknowledge support from The European Communities Seventh Framework Programme (FP7/2007-2013) under grant agreement number ITN 215212 ``Black Hole Universe''. This work was also partially supported by the National Science Foundation under grants AST 0807385 and PHY11-25915 and through TeraGrid resources provided by the Texas Advanced Computing Centre (TACC). We thank SARA Computing and Networking Services (www.sara.nl) for their support in allowing us access to the Computational Cluster. PCF acknowledges support of a High-Performance Computing grant from Oak Ridge Associated Universities/Oak Ridge National Laboratory.

\bibliographystyle{mn2e}
\bibliography{references}{}

\bsp

\label{lastpage}

\end{document}